This is a preprint copy that has been accepted for publication in *Journal of Educational Computing Reasearch*.

Please cite this article as:

# Perceptions of International Female Students Towards E-learning in Resolving High Education and Family Role Strain

Mboni Kibelloh，Yukun Bao[*]

Center for Modern Information Management,

School of Management, Huazhong University of Science and Technology, Wuhan,

P.R.China, 430074

**Abstract**

It is a common phenomenon for many mature female international students enrolled in high education overseas to experience strain from managing conflicting roles of student and family, and difficulties of cross-cultural adjustment. The purpose of this study is to examine perceptions and behavioral intentions of international female students towards e-learning as a tool for resolving overseas high education and family strain from a technology acceptance standpoint. To achieve this goal, Davis's (1989) technology acceptance model is used as the study's conceptual framework, to investigate perceived usefulness, ease of use and behavioral intentions towards e-learning. The research draws on face-to-face interviews with 21 female international students enrolled in classroom taught degree programs at a university in Wuhan, China. The data is analyzed through coding and transcribing. The findings reveal that given its convenience, e-learning is generally perceived as practical in balancing study with family as well as feasible in saving time, money and energy. However, key concerns were raised over the issues of poor and costly Internet connectivity in developing countries, as well as perceived negative reputation, lack of face-to-face interaction and lack of motivation in online environment. Important issues and recommendations are raised for consideration when promoting e-learning programme. This study emphasis on the need to revisit gender supporting policies and effective marketing to re-position the prevailing image of e-learning as a reputable and reliable education delivery method.

[*] Corresponding author: Tel: +86-27-87558579; fax: +86-27-87556437.
Email: yukunbao@hust.edu.cn or y.bao@ieee.org

# 1. Introduction

Strain from managing the conflicting roles of student and family, alongside cross-cultural adjustment difficulties, is a common experience for mature female international student enrolled in high education overseas. Whilst the experience of studying abroad affords learners the opportunity to improve foreign language proficiency, gain greater interest in international affairs, enhance adaptability, personal growth, and competitive advantage in the home employment market (e.g., Carlson and Widaman, 1988; Langley and Breese, 2005), the physical journey to a new country, is often threatened by multiple role demands, institutional barriers, and cross-culture adjustments difficulties formed by perceptions and constraints.

Although literature establishes that stress related to the unavoidable aspect of culture shock, acculturative and adjustment difficulties (i.e. Oberg, 1960) is expected for both men and women, confronted with geographical distance, women particularly mothers face an even greater challenge of nurturing and caring for their children (i.e. Maundeni, 1999). Several scholars including Walkup (2006), identify the particular problems of these students involving: managing academic, childcare and domestic tasks, feelings of exclusion because the tertiary education provider fails or has limited approach to meet specific needs, emotional stress about child-care provision, and guilt in relation to the conflicting roles of "mother" and "student". Compared to other cross-cultural travelers, this cohort of students are unique in that they must not only adapt to new culture and function in an academic setting (Zimmerman, 1995), but must also manage the strain of education and family roles.

Consequently, a large body of knowledge exists relating the emotional ramifications of role strain and cross-cultural adjustment difficulties with various issues such as psychological distress, health alterations, homesickness, the inability to culturally integrate into foreign culture, difficulty learning language, negative academic performance, and high dropout rates (Rindfuss et al. 1980; Stafford, Marian, and Salter, 1980; DeLongis, Folkman, and Lazarus, 1988; Twibell, Ryan and Limbird, 1995; Maundeni, 1999; Allen and Herron, 2003). Though cultural adjustment stressors may vary with time and diminish with duration of stay (Pantelidou and Craig 2006), role strain related stress often remains a persistent concern for women students throughout their academic period. Evidently, these findings necessitate the need to investigate alternative routes of education delivery.

Relevantly, extensive research suggests the use of online distance education (e-learning) as a convenient and flexible education delivery tool for female students. Columbaro (2009) found that e-learning is viable option for busy and working mothers offering education 'anywhere' and 'anytime'. Kibelloh and Bao (2013) showed that online MBA is practical in assisting women enrolled in MBA programmes to balance work, family and career aspirations. Despite the obvious benefits, however, the changing patterns in education have been found to induce anxiety and uncertainty in users (e.g., Ong and Lai 2006; Kibelloh and Bao, 2013), and these concerns have pervasive influence on acceptance of the technology. A better understanding of the perceptions of female international students, would enable institutions to offer suitable programs to meet the needs of prospective online learners. This study aims to examine perceptions and behavioral intentions of international female students towards e-learning as a tool for resolving overseas high education and family strain from a technology acceptance point of view.

To accomplish this goal, Davis's (1989) technology acceptance model (TAM) is used as the most widely employed conceptual framework for predicting acceptance of innovation and technology (i.e. Venkatesh, and Morris, 2000; van Raaij and Schepers, 2008; Pan and Jordan-Marsh, 2010; Sánchez and Hueros, 2010; Bao et al., 2013; Cheung and Vogel, 2013). The TAM model in this study is used to investigate the perceived usefulness, ease of use and behavioral of e-learning in a qualitative case study by interviewing international female students enrolled in classroom taught degree programs in a university in China.

This research makes various contributions to both theory and practice by applying TAM model in a qualitative study investigating international female students perceptions and behavioral intentions towards e-learning as a tool for resolving overseas high education and family strain. Key issues and recommendations are raised for consideration when promoting e-learning programs. Emphases are made on the need to revisit gender supporting policies and effective marketing campaigns to re-position the prevailing image of e-learning as a reputable and reliable education delivery method.

This paper presents a literature review of the barriers related to overseas education based on Cross (1981) conceptual framework. A theoretical account of the TAM theory, e-learning and description of methodology, presents findings leading up to discussion of results and their implications. Recommendations for practice and future research conclude the study.

# 2. Literature Review & Theoretical Framework

## 2.1 Barriers to Overseas Education

Overseas education, referred to by the term *"transnational education",* is defined as "a[A]ll types of higher education study programs, or sets of courses of study, or educational services (including those of distance education) in which the learners are located in a country different from the one where the awarding institution is based. Such programs may belong to the education system of a State different from the State in which it operates, or may operate independently of any national education system." (European Center for Higher Education, United Nations Educational, Scientific, and Cultural Organization, UNESCO-CEPES, 1999). When referring to overseas high education, this UNESCO-CEPES definition is operative.

Accordingly, being enrolled in high education overseas compels international female students to overcome several barriers. Cross (1981) divides these barriers into three categories namely: situational, dispositional, and institutional. The following is a description of each.

### 2.1.1 Situational Barriers

Situational barriers refers to one's situation in life, such as lack of time due to responsibilities on the job or at home, lack of childcare or lack of money, which are likely to deter large numbers of potential learners from education (Cross, 1981; Cross and McCartan, 1984). For example, Maundeni (1999) found that female students, who brought along their children with them to study, experienced financial strain, having to pay nursery or school fees for the children in addition to tuition and living expenses.

Moreover, they were likely to experience time related role strain as a result of looking after the children entirely on their own. On the other hand, Maundeni (1999) also found those who left their children behind greatly missed their children, and believed that their absence adversely affected their studies. Generally, extensive research has shown that childcare is perhaps the most consistent problem faced by mothers who are students (i.e. Griffiths (2002). However, positive support from extended family members, friends and partners, is vital in determining how well mothers cope with these experiences. (i.e. Duncan, 2000; Griffiths, 2002). Overall, time, childcare and financial constrains are situational barriers that can potentially impede on international females' educational goals.

*2.1.2 Dispositional Barriers*

Dispositional barriers are the negative attitudes and self-perceptions about oneself as a learner, which limits student's success (Cross, 1981). Studies have shown that conflict between education pursuits and family often ignites emotional upheaval and stress in the lives of women. Mothers are often caught up in the "Good Mother syndrome" a term used to describe the socially constructed way that mothers are supposed to act and be, mandating women to be completely child-centered and provide active emotional, physical support and entertainment for their children (Hays, 1996; Dillaway and Pare, 2008). The "mother guilt" is a common occurrence for women, who tend to internalize guilt and see themselves as "selfish" for studying (Pare, 2009). Terrell (1990) found that parents usually felt guilt about being unavailable when their children needed them. Carney-Crompton and Tan (2002) argued that depending on the age of their children, those with older children were like to persist to graduation, whereas those with young children were likely to interrupt or stop their education. Lynch (2008) found high attrition rates related to this group of graduate students who faced difficulties combining the two identities of "mother" and "student".

Other studies have highlighted on the effects of the conflict between the role of wife and full-time student. Merrill (1999) discussed the emotional costs for mature women whose partners feel threatened by their participation in higher education and as a result offer their wives very little support. Hughes et al. (1992) found that work and family role strain increases marital tension and decreases marital companionship. Tian (1996) reported on the close relation between divorce rates and women's enrolment in education. Generally, the emotional stress from "mother-child" identity conflict and perceived lack of support from spouse are dispositional barriers that can likely hinder international female students from achieving their educational objectives.

### 2.1.3 Institutional Barriers

Institutional barriers are practices and procedures that exclude or discourage adults from participating in educational activities, due to inconvenient schedules or locations, inflexible school fees, inappropriate course offerings (Cross, 1981). Conversely, various scholars have shown that students often faced difficulties in adjusting to education systems related to particular country. For example, Murphy-Shigematsu (2002) found that students usually had expectations based on the system of their own country or universities in the United States or Europe. The lack of accurate information and pre-arrival orientation impedes the construction of realistic expectations of what students encounter. Furthermore, Deggs (2011) found that new types of institutional barriers related to technology utilization in delivering and enhancing learning existed among adult learners. In addition, colleges and universities tend to lag far behind industry in their tendency to adopt family-friendly policies (i.e. Thompson and Kline, 2000). All in all, Institutions which do not address the unique needs of international students may leave students feeling disappointed, unfulfilled, and even exploited (Shery et al, 2010). These barriers are likely to deter international female students from pursing their educational goals

*2.2 Technology Acceptance Model*

The Technology Acceptance Model (TAM) by Davis (1989) is employed in this study as the theoretical grounding for exploring factors influencing the perception of e-learning. Adapted from the Theory of Reasoned Action (TRA) (Fishbein and Ajzen 1975), the TAM has been used extensively as the theoretical basis for many empirical studies of user technology acceptance (Davis 1989; Venkatesh and Morris 2000; Ong and Lai 2006; van Raaij and Schepers, 2008; Teo and Bahcekapili 2012). The TAM offers scholars a more reliable option to investigate the problem of under-utilized systems and predict behavioral outcomes. Unfortunately, little evidence exists on the application of TAM on international female students enrolled in overseas high education.

The TAM model encompasses two main variables, perceived usefulness and perceived ease of use. Perceived usefulness is defined as an indicator of the extent to which a person believes that using a particular technology will enhance his or her performance and therefore represents an individual's extrinsic motivation to use a technology (Davis 1989). Previous research has established positive effect of perceived usefulness on behavioral intention to use (Davis 1989; Venkatesh and Morris 2000). Conversely, perceived ease of use refers to the degree to which a person believes that the use of a particular technology will be free of effort and is therefore an indicator of an individual's intrinsic motivation to use a technology (Davis 1989). Various scholars (i.e. Venkatesh and Morris, 2000) have shown that a low evaluation of perceived ease of use caused an increase in the salience of such a perception in determining perceived usefulness and user acceptance decisions. According to the TAM model, beliefs that technology is useful and easy to use significantly influences the users' attitudes toward the technology and thereby their decision to adopt the technology. It can be expected therefore, the desire to balance international education pursuit with family responsibilities would likely determine the usefulness and decision to adopt e-learning as a convenient educational route.

*2.3 E-learning*

Increasingly, educators recognize the need to offer e-learning to meet the demands of the students of the twenty-first century (Zhu, 2009). Alternative approaches to international study have emerged to serve the growing need for flexible learning; and many students choose to stay at home while acquiring international education (Hannon & D'Netto, 2007). Accordingly, e-learning enables adults with full-time jobs to attend classes without having to leave their current jobs and allows women to gain high level education while tending to family commitments (Columbaro, 2009; Kibelloh and Bao, 2013). E-learning offers flexible schedule, low travel costs and provides a platform upon which students and teachers from around the world can interact (Hung et al. 2010).

Furthermore, numerous studies have found student satisfaction and perceived usefulness are key factors in explaining learners' behavioral intention to use e-learning (i.e. Liaw 2008; Sánchez and Hueros, 2010; Kibelloh and Bao, 2013). Hurd (2007) discussed on the importance of motivation and self-discipline are necessary for students' success in e-learning. Hrastinski, Keller, & Carlsson (2010) showed that perceived networking, instructor interaction with students influenced e-learning perceptions. Kibelloh and Bao (2013) indicated on the importance of perceived quality in acceptance. Other scholars found perceived ease of use to be an important predictor of behavioral intentions towards online education (i.e. Ong and Lai, 2006; Sánchez and Hueros, 2010). Jashapara and Tai (2011) argued that although people may have different levels of computer experience, innovativeness, and playfulness in e-learning settings, self-efficacy and anxiety play a crucial role in influencing their attitudes towards e-learning. Lack of technological expertise was revealed as the result to fear of working in e-learning environment (Middleton, 2009). Jashapara and Tai (2011) pointed on the critical role of training programs to alleviate computer anxiety.

*2.4 Research Questions*

This study aims to answer the following research questions (RQs):

RQ1. What is the perceived usefulness of e-learning in terms of resolving overseas high education and family strain?

RQ2. What is the perception of e-learning in terms of ease of use?

RQ3. If given the option, would the international female students opt for e-learning?

# 3. Methodology

*3.1 Research Method*

Despite the fact that the TAM has been the subject of investigation in large numbers of research, many of these studies are limited in several respects, including the strictly positivist quantitative approach of research focusing on the adoption of technologies (e.g.,

van Raaij and Schepers, 2008; Pan and Jordan-Marsh, 2010; Sánchez and Hueros, 2010; Cheung and Vogel, 2013). This decision is in accordance with recommendations of proponents of the case study approach, such as Yin (1994). Moreover, the voices of individuals as potential adopters of innovation have not been prominent in technology acceptance investigations. This study helps address this limitation in the literature by providing an in-depth qualitative study by recording participants views of e-learning to obtain rich and descriptive narratives rather than using structured questionnaires and risk missing important information. Accordingly, Merriam (1998, p. 17) advises that qualitative research is "A highly subjective phenomenon in need of interpreting rather than measuring". This method will provide a foundation upon which other researchers can build in generating hypotheses to be tested in quantitative designs (Silverman, 2006).

*3.2 Participants*

Face-to-face interviews were conducted in the spring of 2013 with 21 international female students enrolled in classroom taught programs at a University in Wuhan, China. In this study, international students, referred to by the term *"internationally mobile students"*, are students who are noncitizens of the host country, who do not have permanent residency, and who did not complete their entry qualification to their current level of study in the host country (United Nations Educational, Scientific and Cultural Organization, UNESCO, 2006). The participants came from Thailand (four participants), Zambia (two participants), Tanzania (two participants), Zimbabwe (one participant), Iraq (two participants), Rwanda (one participant), Pakistan (three participants), Columbia (one participant), Nigeria (one participant), Vietnam (two participants), Cambodia (one participant), and Uzbekistan (one participant). The average age of the participants was 28 years old. There academic levels included Bachelor (four participants), master's (nine participants) and PhD (eight participants). Seven were married, while fourteen were single. From those married, five had children with them in Wuhan China; one had a child left in her home country.

This study employed the snowball sampling method to recruit the international female students. This technique, which relies on referrals from the initial subjects to generate additional subjects, was suitable and effective for casting a wide net over variety of participants from different countries. The participants were selected using the first author's primary classmate contacts and network selection. Moreover, to diversify the sample participants were solicited from various sites of the community (i.e. international student hostels and classrooms and labs). The personal contacts were then encouraged to invite individuals with similar backgrounds, international female students who were interested in taking part in the interview.

*3.3 Interview protocol and data collection*

Interview questions were given to participants prior to the interview, and participants were guaranteed the confidentiality of their participants to ensure they spoke freely. First, as recommended by Merriam (1998), the protocol began with questions designed to gather demographic data, age, e-learning experience, academic background, and major. Next, the interviewer asked for open-ended questions concerning the participant's perception of a) the usefulness of e-learning, b) the challenges of e-learning c) the perceived ease of use of e-learning, and d) whether they would take-up e-learning if given the option. The questions were in line with the guiding research questions that stemmed from the TAM theoretical framework (Davis, 1989).

The interviewers audiotaped the interviews, and took notes to ensure accurate recording of the responses and the interviewer's overall impressions. To protect the participants' identities, pseudonyms are used in this report. All interviews were conducted in English and generally lasted one hour each. The data were then sorted into a database manually by one of the authors and checked by the other author.

*3.4 Data analysis*

Following Braun and Clarke's (2006) qualitative data analysis model, this study adopted the theoretical thematic data analysis to analyse the case data. The analysis required several steps. First, the researchers read through the transcripts and jotted down comments, notes, thoughts, and observations in the margins. Next, the researchers summarized the marginal notes by sectioning, grouping similar data into categories. Code labels were assigned to each section using the interviewee's or the researchers' words. The preliminary codes were examined for overlap and redundancy. The authors then eliminated redundant codes and collapsing similar codes, which enabled the codes to be constructed in the early stage and to be narrowed down to broader themes. The new list of code words was then used to examine the texts to check whether these codes revealed common themes and recurring patterns. The different data sets were continuously read and analysed to refine the categories and to ensure that no text sections were overlooked.

While analysing, the researcher continuously linked the recurring themes to the TAM as the theoretical lens of this study. The TAM was used to organise the categories and form in-depth views of the conceptual meanings of the categories under the framework. The themes were grouped into three main categories including, perceived usefulness, challenges, ease of use, behavioural intentions. The different components of the TAM model (i.e., perceived usefulness and ease of use) were used as ''containers'' for arranging data themes (Barab et al. 2003).

As the interviewers gathered more data and coding continued, it was found that no new themes were being identified and no new issues arose. Therefore, as suggested by Strauss and Corbin (1990), this study had reached its saturation point with 21 interviews. To protect participants' identities, pseudonyms are used in this report.

To ensure trustworthiness of study, both researchers were involved in the analysis. Where necessary, the participants were contacted for clarification or additional information. Peer debriefing was also employed. A professional peer who was not directly involved in the data collection but was familiar with the socio-cultural aspects of qualitative case study analysis was invited to comment on the findings as they emerged and to check for misinterpretations and researcher bias. Furthermore, the participants were presented with the findings to assess the extent of their agreement with the provided interpretations.

## 4. Findings

### 4.1 Balance family and study

Most of the participants believed that e-learning was indeed useful for students with children in terms of time management, allowing mothers to allocate the time for family and studies accordingly. They expressed:

> Personally e-learning is a very useful kind of learning in terms of time management. As lady I have the time to look after my family and work at the same time whilst learning. It is easily accessible without making an appointment without a lecturer. Ester, 28, Bachelor student, Zambia, married, 1 Child with her

I have always wanted to have online learning. E-learning can help me to achieve my academic goal as well as staying with my family as personal life goal. Fiona, 24, Cambodia, Married, no children

Incase if anyone has a family or responsibility can consider e-learning as very useful instead of travelling overseas - Amina, 35, PhD student, Iraq, Married, 1 Child with her

## 4.2 Flexibility and convenience

Most of the participants believed the e-learning was useful for those who cannot not attend university due various economical or physical constraints. Typical examples include:

I think e-learning is very useful. If one is knowledge lover but cant get time to go to class/campus that person can get higher or professional education easily - Poonam, 33, Masters student, Pakistan, married, 2 children with her

(with e-learning), I can repeat the course whenever I want to (learning at my own pace). - Fiona, 24, Masters student, Cambodia, Married, no children

I think is a useful tool to give education access to people that cannot go to university for economic or other reasons. Is useful in that it can be used in anywhere at any time and can create a lot of connection between people from many different parts of the world. - Gloria – 29, PhD student, Columbia, single, no children

## 4.3 Readily available course materials

Few participants believed given the wealth of knowledge repository online, e-learning was resourceful in offering wide variety of course materials online. They mentioned:

> Course/study materials are always readily available - Margaret, 28, PhD student, Public Health Nigeria, single, no children

> Information is easily gotten from the Internet, making it very useful. - Ester, 28, Bachelor student, Zambia, married, 1 child with her here

### 4.4. Costly vs. cost effective

Another theme that arose from the interviews was the conflicting perceptions of e-learning cost benefit. While some reckoned e-learning was cost effective in terms of significantly cutting down tuition fees, travel fares and living allowance, others believed that e-learning was expensive given the high internet costs in some countries.

> It (e-learning) can be cost effective. Even a middle class person (male /female) can get education. For example in developing countries women want to study more but financial and social barrier can't let them do this. Moreover, if e-learning includes programs for working class like farmers, welder…its good. It can help in increase literacy rate.- Poonam, 33, Masters student, Pakistan, married, 2 children, with her

> I think it is not easy for everybody to use e-learning, because in some areas of a country Internet is expensive - Grace, 23, Masters student, Uzbekistan, single, no children

### 4.5 Reliability and reputation

A number of the participants expressed skepticism towards the reliability and reputation of e-learning. Few argued that in their countries, e-learning was not recognized as a legitimate method of education delivery. This is based on the assumption that online education lacks adequate means of testing student knowledge gain and lacks teacher interaction, which is better source information.

> It is still difficult to really know if you are signing up for a serious educational program, in terms of payment, and consistence for the students to follow the whole course until the end. - Gloria – 29, PhD student, Columbia, single, no children

> E-learning must become popular to be accepted by the world (some universities and companies). So it can be easy for you to present your degree anywhere - Veronica, 28, Masters student Rwanda, single, no children

## 4.6 Ease of use

Generally, the participants found e-learning effortless to use. However, few participants noted that e-learning would only be easy if student was already well versed with computer and Internet skills.

> E-learning has become frequently used in the educational system all around the world and I think it is very easy to use it because we live in renaissance age and no one cannot not know how to use computer.- Sharon, 39, PhD student, Iraq, Single, no children

> I think e-learning is not easy to use making it so difficult for anyone without or with limited computer background to use it. - Marilyn, 27, Masters student, , Zimbabwe, Single, no children

> Personally, it is easy as I am quite exposed to IT stuff and also Internet access. - Fiona, 24, Master, School of Management, Cambodia, married, no children

## 4.7 Poor Infrastructure & Access

Many of the participants expressed much concern over the availability of stable Internet to support e-learning in respective countries. Although many participants from the developing world found e-learning usefulness, many wondered how the poor villages and towns with unreliable electricity, uneven distribution of Internet coverage could sustain e-learning and expensive Internet charges. They stated:

> There are a lot of challenges with e-learning especially in African countries. One of them is low Internet speed. This is a huge problem especially in my country, where we can experience much delay in accessing materials from Internet. Moreover there are some places where people cannot access these online courses, resulting from shortage of technology especially in the countryside (villages) - Maryam, 23, Bachelor student, Tanzania, single, no children

> Access to Internet, the majority in my country rather Africa at large have no access to Internet. If it happens that you have access to internet back in Africa, it happens that you can not sit for a long time browsing for it is so costly and very slow hence your limitation in finishing your assignments. - Ester, 28, Bachelor student, Zambia, married, 1 Child with her

> In Cambodia, the internet access somehow is still not good in terms of speed so it is not favorable for live e-learning. - Fiona, 24, Master student Cambodia, Married, no children

### 4.8 Inappropriate for some courses

In addition, other participants particularly those from science, technology, engineering and mathematics (STEM) perceived e-learning as inappropriate for their academic majors.

> It is not suitable for some health-related programs where clinical study is required. Margaret, 28, PhD student, single, no children

Although e-learning has many benefits for students, it also is limited to some subjects while other majors may not be appropriate for e-learning. - Sharon, 39, PhD student, Iraq, single, no children

### 4.9 Lack of motivation

Few students indicated the importance of motivation required in self and online study especially given the many distractions found online and on electronic gadgets. The participant shared:

Unless someone really has the zeal to take it and make it work, e-learning can be almost useless. Nowadays with the many applications available it's easy to end up doing something else more interesting on your phone or computer such that face-to-face supervision especially for younger generation is still essential. - Marilyn, 27, Masters student, Zimbabwe, Single, no children

Sometimes it can make people get bored because e-learning program cant participate and answer the question immediately. Sophia, 24, PhD student, Thailand, single, no children

### 4.10 Lack of interaction & Delayed Response

Although several participants found that e-learning was useful, they argued that it can be a lonely experience given the lack of interaction with instructor. Moreover, delayed feedback was also seen as a disadvantage.

e-learning has delays in responses compared to face-to-face class discussions and also gives a feeling of loneness and isolation.. - Khadija, 30, Bachelor student, Tanzanian, Married, 2 children with her

But when you study by e-learning you cant discuss with other people or cant suddenly ask the teacher a question. But it is confortable for your life. - Sara, 23, Masters, Enterprise management, Thailand, Single e-learning in a way cannot respond immediately to what I want to know if its not live learning. - Fiona, 24, Master, School of Management, Cambodia, Married, no children

*4.11 Health reasons*

Other participants were concerned about the health implications of engaging in e-learning, and argued that spending too much time in front of the computer was unhealthy. They remarked:

I don't like exposing myself to electronic device for long time as it hurts my eyes and stresses me out. - Fiona, 24, Master, student, Married, no children

I think it has limitations to user in terms of danger to health - Grace, 23, Masters student, Uzbekistan, single, no children

## 5. Discussion & Implications

The findings of this study present several important issues for discussion. The following section discusses these results as per the research objectives:

*5.1 RQ1. What is the perceived usefulness of e-learning in terms of resolving overseas high education and family strain?*

Overall, the participants valued the benefits of e-learning in terms of flexibility and convenience, based on the prospect of pursing international education without having to travel abroad, while allowing for active involvement in family and work affairs. This finding resonates with Columbaro (2009) who suggests that e-learning is useful for learners needing to balance multiple roles. Moreover, the participant credited online learning for fostering students' communication and collaboration with teachers, and students from around the world. Additionally, e-learning was perceived as useful in resolving constrains such time, money and other challenges that are met by international students while adjusting into foreign countries. Future research can examine educational outcomes (i.e. learning performance, or dropout rates) related to perceived gains and satisfaction of students enrolled in e-learning programmes while managing family demands.

One the other hand, despite the perceived benefits of e-learning, a concern that emerged from the findings, particularly by students from developing countries was the issue of poor communication infrastructure in their home countries. They explained that the unequal distribution of Internet coverage particularly in the remote villages was an impediment to e-learning. This finding supports the study of Folorunso et.al (2006). Furthermore, participants were concerned about the general reputation of e-learning especially in regards to their home country's employment market. These findings suggest that more research attention needs to be focused on ensuring quality and effective means of marketing the various benefits of online education, especially while considering different cultures. Online educational providers must also be attentive to the needs of existing e-learners so as to maintain a continuous e-learner base and benefit from informal marketing through positive word of mouth.

Furthermore, other students raised concerns over the lack of networking and interaction in online classes. This echoes Chu (2008) who argued that interaction and active discussion are key success factors for learning effectiveness in Internet-based distance education. Given the geographical distance, establishing a healthy tone or climate is of utmost importance. Instructors could consider using Learning Management

Systems (LMS) to facilitate for the interaction between instructors and students, by detail tracking students' progress, and delivery of content through the web (Lonn and Teasley 2009). In addition, Hrastinski, Keller, and Carlsson (2010) suggested using synchronous conferencing such as skype, that resemble face-to-face interaction, to allow for more direct social interactive and feedback among learners and teachers. This would also mitigate motivational issues that were raised by a few participants. Additionally, Starr-Glass (2013) suggests enhancing e-learner motivation by fostering mentoring relationships at a distance where mentors meet with their students during several annual meetings and emphasizes that mentors require having a familiarity of distance learning in general and understand mentoring dynamics in order to make e-mentoring effective. This study also supported Şad and Özhan (2012) findings related to health issues, where students complained about watering and irritation in eyes as a result of looking at their screens for a long time.

### 5.2 RQ2. What are the perceptions of e-learning in terms of ease of use?

Majority of participants believed e-learning was somewhat easy to use. They indicated that a big part of their daily academic affairs i.e. through course assignment, desk research, PowerPoint presentations, or Chinese online language demanded them to use computers and the Internet. Moreover, the Internet was used as the predominant means of communicating with family back home through voice over IP applications such as Skype. Regardless of the perceived ease of use, the participants emphasized that using e-learning required a particular set of skills and that only those proficient with computer usage would enjoy the benefits. This resonates with Shana (2009) argument that students in an online environment need to have attained a certain computer competence level.

Jashapara and Tai's (2011) findings show that e-learning systems need to adopt special measures for individuals with low levels of confidence with IT and high levels of anxiety. Institutions could consider prerequisite starter-up training course, prior to student enrolment irrespective of individual's computer experience and background. To serve an array of international students, providers would need to establish a full-time dedicated online foreign student advisory and helpdesk to assist in foreign e-learners endeavors.

*5.3 Q3. If given the option would the international female students opt for e-learning?*

The participants' behavioral intentions towards e-learning reflected mixed feelings. Although the participants had never been previously enrolled in e-learning degree programs, majority revealed that they could opt for e-learning in lieu of traveling overseas to spend time with family as well as save money, time and energy. For example, Ester from Zambia believed that e-learning would allow her more time to bond with her child. Margaret from Nigeria expressed that online education would afford her the opportunity to further her education while pursuing other life goals. Others stated that they could consider adopting e-learning as long as the Internet was stable. A few students saw e-learning as a good opportunity to attend virtual lectures by well renowned lecturers, and a great platform to share ideas with other learners worldwide.

On the other hand, several participants expressed the significance of travelling for education. For instance, Fiona from Cambodia wanted to experience life overseas and preferred physical class environments over e-learning. Similarly, Gloria from Columbia believed that e-learning was limiting in terms of gaining cultural exposure of a host country as well as of fellow international students. These findings support Salisbury et al. (2009) in that students go abroad to enhance their education through international experience. Online education practitioners might want to factor in creative ways in which to promote host country's culture as well as nurture cultural exchanges among online international students. One way to achieve this is through short-term study or campus visits abroad (Holland and Kedia, 2003), for cultural and language experiences, as well as for interacting and building relationships with professors fellow students.

Other participants reckoned that e-learning was not suitable for their academic disciplines. For example, Marilyn from Zimbabwe believed that given difficultly to simulate lab work in virtual classroom, e-learning would be inappropriate for those whose departments are based in the lab. Moreover, Marilyn felt that in the technical jargons and language barrier already hampered face-to-face discussions in the scientific fields, therefore, e-learning would not necessarily be helpful. Furthermore, few participants showed no interest in e-learning based on perceived reputation. For instance, Amina from Iraq mentioned that e-learning was not recognized as quality education in her country, as people in Iraq still believe that good education involves the physical presence of a teacher and examinations.

*5.5 Limitation*

This study has the following limitation: While the use of the snowballing sampling method proved most suitable for this study, it introduced bias, as the method reduces the likelihood that the sample will represent a good cross- section of the population of international students. This shortcoming was demonstrated in this research by the predominance of students from developing or transitional economies. Perhaps this can be explained by fact that these countries represent the largest population of foreign students in China. Moreover, China's Ministry of Education offers a number of scholarships to these countries as in accordance to the China's foreign policy for strengthening international relations. Thus, these findings may not necessarily be applicable to women from predominately western cultures such as England or US. In addition, the women in this study are not a homogeneous group. Their exists significant differences among the participants in terms of the degree of individualism and collectivism which can assumedly affect the level of accountability and affinity towards family responsibilities as well as influence individual cultural adjustment process.

## 6. Conclusion

As labor market demands shift, mature female students seeking to upgrade, refresh or complement their knowledge, are increasingly entering universities (i.e. Maundeni, 1999; Yelland, 2011); and many are choosing to study outside their country borders (Chang, 2012). However, the hassles of balancing multiple roles together with institutional barriers formed by perception and constraint often hinder these female students from attaining an overseas higher education. These complications hamper women from completing their education, and arguably represent an early stage of a situation characterized as a 'leaky pipeline', where women (and especially mothers), more likely than men drop out of the 'pipeline' of established career trajectories (Dugger 2001).

Prompted by the growing concerns of international female students welfare, the current study contributes to the endeavor by examining the alternative educational delivery method of e-learning and subsequent behavioral intentions. Though the overall findings of this research reveal a positive outlook towards its usefulness particularly in education-family conflict resolution, traditional views of classroom learning and teaching still have a strong hold on the educational values of many learners. Still, more light needs to be shed on the potential of e-learning. However, it can be hoped that the e-learning movement led by the early adopters "innovators" and advocating champions, can help shifting the consciousness of the general public toward its acceptance. How well these e-learning champions manage their roles (i.e. as e-learners and caregivers) could effectively change not only in their families but also interest other mothers who may be contemplating further education. E-learning awaits future research on ways to effectively market its benefits and re-position its image as a trustable and reliable education delivery channel. This study provides a platform to advance future knowledge, theory and practice.